%% file: main.tex
\documentclass[chi_draft]{sigchi}

\CopyrightYear{2020}
\setcopyright{acmlicensed}
\doi{https://doi.org/10.1145/3313831.XXXXXXX}
\isbn{978-1-4503-6708-0/20/04}
\conferenceinfo{CHI'20,}{April  25--30, 2020, Honolulu, HI, USA}
\acmPrice{\$15.00}

\usepackage{balance}       
\usepackage{graphics}      
\usepackage[T1]{fontenc}   
\usepackage{txfonts}
\usepackage{mathptmx}
\usepackage[pdflang={en-US},pdftex]{hyperref}
\usepackage{booktabs}
\usepackage{textcomp}
\usepackage{color,soul}
\usepackage{multirow}
\usepackage[utf8]{inputenc}

\PassOptionsToPackage{hyphens}{url}\usepackage{hyperref}

\usepackage{caption}
\usepackage{subcaption}
\usepackage{float}
\usepackage{graphicx}

\newcommand{\ignore}[1]{}

\newcommand{\mSebisName}{SSBS}
\newcommand{\subscaleSocial}{Social}
\newcommand{\subscaleTech}{Tech}

\usepackage{microtype}        
\usepackage{ccicons}          

\usepackage{todonotes}

\def\plaintitle{Smartphone Security Behavioral Scale: A New Psychometric Measurement for Smartphone Security}

\def\emptyauthor{}
\def\plainkeywords{Authors' choice; of terms; separated; by
  semicolons; include commas, within terms only; this section is required.}

\makeatletter
\def\url@leostyle{%
  \@ifundefined{selectfont}{
    \def\UrlFont{\sf}
  }{
    \def\UrlFont{\small\bf\ttfamily}
  }}
\makeatother
\def\pprw{8.5in}
\def\pprh{11in}

\definecolor{linkColor}{RGB}{6,125,233}
\hypersetup{%
  pdftitle={\plaintitle},
  pdfauthor={\emptyauthor},
  pdfkeywords={\plainkeywords},
  pdfdisplaydoctitle=true, 
  bookmarksnumbered,
  pdfstartview={FitH},
  colorlinks,
  citecolor=black,
  filecolor=black,
  linkcolor=black,
  urlcolor=linkColor,
  breaklinks=true,
  hypertexnames=false
}
\usepackage[breakable, theorems, skins]{tcolorbox}
\tcbset{enhanced}

\makeatletter
\def\@copyrightspace{\relax}
\makeatother
\begin{document}

\title{\plaintitle}

\numberofauthors{6}

\author{%
  \alignauthor{Hsiao-Ying Huang\thanks{The authors contributed equally.}\\
    \affaddr{Illinois Informatics Institute,}\\
    \affaddr{University of Illinois at Urbana-Champaign}\\
    \email{hhuang65@illinois.edu}}\\
  \alignauthor{Soteris Demetriou$^*$\\
    \affaddr{Department of Computing,}\\
    \affaddr{Imperial College London}\\
    \email{s.demetriou@imperial.ac.uk}}\\
  \alignauthor{Rini Banerjee\\
    \affaddr{Department of Computing,}\\
    \affaddr{Imperial College London}\\
    \email{rini.banerjee18@imperial.ac.uk}}\\
  \alignauthor{Güliz Seray Tuncay\\
    \affaddr{Computer Science,}\\
    \affaddr{University of Illinois at Urbana-Champaign}\\
    \email{tuncay2@illinois.edu}}\\
  \alignauthor{Carl A. Gunter\\
    \affaddr{Computer Science}\\
    \affaddr{University of Illinois at Urbana-Champaign}\\
    \email{cgunter@illinois.edu}}\\
  \alignauthor{Masooda Bashir\\
    \affaddr{School of Information Sciences,}\\
    \affaddr{University of Illinois at Urbana-Champaign}\\
    \email{mnb@illinois.edu}}\\
}


\maketitle

\begin{abstract}
Despite widespread use of smartphones, there is no measurement standard targeted at smartphone security behaviors. In this paper we translate a well-known cybersecurity behavioral scale into the smartphone domain and show that we can improve on this translation by following an established psychometrics approach surveying 1011 participants. We design a new 14-item \textit{Smartphone Security Behavioral Scale (SSBS)} exhibiting high reliability and good fit to a two-component behavioural model based on technical versus social protection strategies. 

We then demonstrate how SSBS can be applied to measure the influence of mental health issues on smartphone security behavior intentions. We found significant correlations that predict SSBS profiles from three types of MHIs. Conversely, we are able to predict presence of MHIs using SSBS profiles. We obtain prediction AUCs of 72.1\% for Internet addiction, 75.8\% for depression and 66.2\% for insomnia.

\end{abstract}


\ignore{

}


\section{Introduction}
\label{sec:introduction}
\input{introduction.tex}

\section{Related Work}
\label{sec:related_work}
\input{related_work.tex}

\section{Approach of Psychometrics}
\label{sec:psych_approach}
\input{psy_approach.tex}


\section{Phase-1 Study: Building the scale upon SeBIS}
\label{sec:phase1}
\input{phase_1_study.tex}

\section{Phase-2 Study: Developing SSBS}
\label{sec:phase2}
\input{phase_2_study.tex}

\section{Applying the Scale: relation to mental health}
\label{sec:mhi_method}

\input{mhi_method.tex}
\section{Correlations between Mental Health Issues and Smartphone Security Behavior Intentions}
\label{sec:correlations}
\input{correlations.tex}

\section{Predicting Smartphone Security Attitudes}
\label{sec:predicting_sec}
\input{predicting_sec.tex}

\section{Predicting Mental Health Issues}
\label{sec:predicting_mhi}
\input{predicting_mhi.tex}

\section{Discussion \& Future Work}
\label{sec:discussion}
\input{discussion.tex}

\section{Conclusion}
\label{sec:conclusion}
\input{conclusion.tex}


%
%
%
%
%
\balance{}

\bibliographystyle{SIGCHI-Reference-Format}
\bibliography{sample}

\end{document}

%% file: introduction.tex
Smartphones have become an essential part of modern society. In 2018, about 77\% of U.S. adults owned smartphones, an increase of 42\% since 2011~\cite{pew}. Internationally, there are 3.3 billion active smartphone users, accounting for about 43\% of the whole world population~\cite{kooistra2018}. With the advancement of mobile technologies, smartphones are now involved in almost any daily activity. For instance,
70\% of smartphone users exchanged emails, 65\% connected with their friends and families on social media, 73\% looked up for directions, 59\% watched videos, and 45\% did online shopping ~\cite{Adobe2017report}. However, as smartphones have become a hub for storing and accessing personal sensitive information~\cite{kang2011usage}, an increasing number and diversity of malicious parties aim to exploit security vulnerabilities of smartphones and their users.

Mobile operating system developers have been dedicated to equipping their operating systems with a number of counter measures (discretionary and mandatory access control, trusted computing \textit{etc.}). However, the security of such systems still heavily relies on the behaviors and decision-making of users. For instance, Android and iOS feature a permission model to enable users to decide if they want to grant application  requests to access sensitive system resources and information. Some repackaged malware apps aim to trick users by mimicking the look-and-feel of popular legitimate apps with subtle differences in their title or logo to attract users to download and trust the app and further grant the permissions~\cite{zhou2012detecting}. Other attacks target intricate configuration properties of smartphones: attackers can extract users' passwords when they access sensitive web domains (\textit{e.g.}~their bank account) from their smartphones on a public network~\cite{li2016csi}; users who never reset their advertising ID, can be subjected to fine-grained profiling by advertising libraries; users who are not attentive to the information provided by websites or applications can become the victims of phishing attacks~\cite{muppavaram2018safe}. Since humans play a such critical role in smartphone security, it is important to understand users' smartphone security behaviors.

When it comes to user behavior on smartphones, previous studies have investigated users' perceptions, attitudes, and behaviors toward smartphone security. They found that users tend to ignore warnings/messages~\cite{felt2012android, kelley2013privacy}, have the misconceptions of the operation of smartphone security~\cite{kelley2013privacy, mylonas2013delegate}, and show minimal attempts to protect their smartphones~\cite{das2016security, mylonas2013delegate}. Users' careless behavior on smartphones can result from their misunderstanding of smartphone operations. Most smartphone users view their smartphone as just a mobile device for entertainment and communication; they are not conscious that their device is actually a handheld computer that is vulnerable to a wide range of cyber attacks~\cite{kulkarni2015vulnerabilities}. In these circumstances, will users have different security protection behaviors from how they address security on desktop and laptop computers? If so, how can we measure users' smartphone security behavior in a systematic way across contexts?

Prior studies have measured users' smartphone security behaviors in different ways. Some studies adopted field observation, and some of them employed a self-reported approach~\cite{das2016security, jones2015efficacy, thompson2017security}. Since  field observations usually require more resources and have limitations in assessing full aspects of security behaviors, most studies utilized self-reported measurements. In terms of self-reported measurements, we found that many of them developed their own measurements based on computer security or adopted from  smartphone measurements in other contexts. Therefore, based on the current literature, there is an need for a measurement system for security behavior that is standardized and specific to smartphones.

To fill this research gap, the goal of this study is to provide a model and analysis to support a standardized scale for measuring users' smartphone security behavior intentions based on a systematic psychometric approach~\cite{netemeyer2003scaling}. We present a study with two phases evaluated on a total of 1011 participants. In our phase-1 study, we examined if the model of general computer security behavior intentions could be applied to smartphone security behavior intentions. We adopted four dimensions from a well-established measurement, Security Behavior Intention Scale (SeBIS), developed by Egelman and Peer~\cite{egelman2015scaling} and examined the fitness of these dimensions on users' smartphone security behavior by factor analysis. Our findings indicate that \textit{smartphone security behavior intentions entails new dimensions that are different from the model of general computer security behavior intentions}. Therefore, in our phase-2 study, we \textit{created a new scale measurement for smartphone security using systematic scale development procedure}. We also used multiple statistical analyses to ensure the reliability and validity of our scale.

We illustrate potential applications of this measurement method by applying our new scale to study the effects of mental health issues (MHIs) in smartphone security behavior intentions. This was partly inspired by recent findings suggesting that smartphone \textit{usage} can correlate with MHIs~\cite{perez2019Cyberphychology}. Our main objective is to answer the following understudied research question: \textit{Do smartphone security behavior intentions correlate with mental health issues?}  Toward this, we analyze the relationship between  smartphone-specific security behavior intentions with prevalent MHIs such as Depression, Insomnia and Internet Addiction. We leverage our online survey to find that, there exist strong correlations between smartphone security behavior intentions and these MHIs. To explore further, we performed a statistical analysis to study the predictability of smartphone security behavior intentions from mental health scores. Here are two examples of our findings. First, we found that participants with a score indicating presence of depression are \textit{more likely} to perform stringent security technical configurations on their smartphones in comparison with participants with lower depression scores. Second, we found that participants with a score indicating no presence of severe Internet addiction are \textit{more likely} to take stringent social security decisions on their smartphones in comparison with participants with higher Internet addiction scores.


To further explore the relationship between smartphone security behavior intentions and mental health indicators, we then asked the following research question: \textit{Can security behavior intentions be used as predictors of MHIs?} Even for a modest size of data set, we indeed found strong evidence that off-the-shelf binary machine learning classifiers trained with security behavior intentions, can be leveraged to effectively predict an MHI in a user based on his or her security behavior on a smartphone. We also performed experiments to identify how important is each SSBS behavior in those prediction tasks and demonstrate how these behaviors can be monitored in practice on Android, the most widely used smartphone operating system.



Our contributions are summarized as follows:

\noindent$\bullet$ We found that new dimensions are important when measuring smartphone security behaviors. These are different from the general security behavior model.

\noindent$\bullet$ We developed a new standardized scale for smartphone security behavior intentions (SSBS) which is based on two factors (technical behaviors versus social behaviors) and showed good psychometric properties with high internal consistency.

\noindent$\bullet$ We are the first to explore the effect of MHIs in smartphone security behaviors.

\noindent$\bullet$ We found significant correlations between smartphone security behaviors and some common MHIs.

\noindent$\bullet$ We developed feature sets based on SSBS that can be used in machine learning models to predict  certain MHIs. We analyzed the importance of each feature on those prediction tasks and demonstrated how they can be monitored in practice.

\ignore{
\begin{itemize}
    \item  We found that new dimensions are important when measuring smartphone security behaviors. These are different from the general security behavior model.
    \item We developed a new standardized scale for smartphone security behavior intentions (SSBS) which is based on two factors (technical behaviors versus social behaviors) and showed good psychometric properties with high internal consistency.
    \item We are the first to explore the effect of MHIs in smartphone security behaviors.
    \item We found significant correlation between smartphone security behaviors and some common MHIs.
    \item We developed feature sets based on SSBS that can be used in machine learning models to predict  certain MHIs. We analyzed the importance of each feature on those prediction tasks and demonstrated how they can be monitored in practice.
\end{itemize}
}
In the next section, we will review the literature that are most related to our study and highlight research questions based on our observed research gaps.


%% file: related_work.tex
\ignore{Smartphone security has been researched from different aspects. More related to our work are studies investigating smartphone security from the perspective of users.}Prior works have examined users' perceptions, attitudes, and behaviors toward smartphone security. Their findings can be summarized into three realms: the inattentiveness toward security warnings/messages~\cite{felt2012android, kelley2013privacy}, the misconceptions of smartphone security~\cite{kelley2013privacy, mylonas2013delegate, chin2012measuring}, and low level of smartphone security behavior~\cite{das2016security, mylonas2013delegate}. 





In terms of behavioral measurements, previous studies have assessed users' smartphone security behaviors via two main approaches: field observation and self-reported measurement. For field observations, most studies focus on two aspects: users' authentication and locking behaviors on smartphones~\cite{harbach2014sa, harbach2016anatomy, gascon2014continuous}, and users' behaviors on granting access~\cite{fisher2012short, wijesekera2015android, almuhimedi2015your}. Although field observation can probe into users' actual behaviors in the real world, it usually focuses on a single aspect of the behavior, making it difficult to conveniently gain a comprehensive understanding on users' behaviors in a short period of time. Therefore, many studies adopt a self-reported approach to measure users' smartphone security behaviors. 

There are various means to measure self-reported smartphone security behaviors. The most used approach in prior work is to develop measurements by adapting more general computer security assessments or by modifying a developed measurement from previous studies. For example, Das and Khan~\cite{das2016security} generated a 6-item measure that was adapted from Microsoft's computing safety index. Jones and Chin~\cite{jones2015efficacy} performed a survey study to investigate students' usage and security behaviors on smartphones by asking seven questions about security practices. A more recent study by Thompson \textit{et al.}~\cite{thompson2017security} designed a five item measure to assess smartphone security behavior in a personal context, which was adapted from a security behavioral assessment in personal computer usage by Liang and Xue~\cite{liang2010understanding}. Another recent (2018) survey study by Verkijika~\cite{verkijika2018understanding} examined south African users' smartphone security practices by using five questions that were adapted from the measurement developed by Thompson et al.~\cite{thompson2017security}. 

While reviewing developed security measurements\ignore{(see Table~\ref{table1})}, we found that there is no standardized and targeted way to measure smartphone security behavior (or behavior intentions) across different contexts. Existing methods are all adopted or adapted from general computer security behavior. However, it is possible that users' smartphone behaviors can deviate from their computer behaviors. For instance, Chin et al.~\cite{chin2012measuring} found that participants' behaviors and activities on smartphones were quite different from their use of laptops. For example, users were less likely to purchase and perform sensitive tasks on their smartphones because of security concerns regarding mobile devices. 

We Identify two key gaps in the current literature: 1) there is no standardized measurement of smartphone security behaviors across contexts; 2) it remains unclear if general computer security behavior measurements is adequate to apply to assess smartphone security behaviors. Thus, a key goal of this study is to develop a standardized and valid measurement of smartphone security behavior intentions that can be used in different contexts. Toward this goal, we ask the following concrete research questions: 
\begin{itemize}
\item \textbf{RQ1:} How adequate is the application of general computer security behavior measurement to smartphone security?
\item \textbf{RQ2:} If the application of computer security behavior is not adequate, can we develop a measurement tool which can capture smartphone security behavior intentions?
\end{itemize}

%% file: psy_approach.tex
\begin{table*}[!ht]
\centering
\caption{Preliminary set of survey items developed based on SeBIS (\textit{smartphone-SeBIS})}
\label{table2}
\resizebox{\textwidth}{!}{%
\begin{tabular}{lllll}
Dim. & ID & Item & $\mu$  & $\sigma$ \\
\hline
\multirow{5}{*}{Device Securement} & DS1 & I use biometrics (fingerprint, face recognition) to unlock my smartphone. & 2.41 & 1.6 \\
 & DS2 & I enable encrypted storage on my smartphone. & 2.41 & 1.44 \\
 & DS3 & I use a rooted/jailbroken phone (r). & 1.45 & 1.07 \\
 & DS4 & I turn on the “lost my device” feature on my smartphone. & 2.5 & 1.6 \\
 & DS5 & I use a password/passcode to unlock my smartphone. & 3.76 & 1.51 \\
\hline
\multirow{3}{*}{Password management} & PM1 & I regularly change my password for online services/accounts using my smartphone. & 2.36 & 1.13 \\
 & PM2 & I share my smartphone's passcode/PIN with other(s). (r) & 1.51 & 0.99 \\
 & PM3 & I use password manager app to manage my passwords on my smartphone. & 1.88 & 1.31 \\
\hline
\multirow{9}{*}{Proactive awareness} & PA1 & When downloading an app, I check that the app is from the official/expected source. & 3.95 & 0.99 \\
 & PA2 & Before downloading a smartphone app I ensure the download is from official application stores (e.g. Apple App Store, GooglePlay, Amazon Appstore) & 4.13 & 1.14 \\
 & PA3 & I reset my Advertising ID on my smartphone. & 1.6 & 1.02 \\
 & PA4 & I manually revoke permissions from apps. & 3 & 1.09 \\
 & PA5 & I grant smartphone apps the permissions they request. (r) & 3.2 & 0.85 \\
 & PA6 & I disable geotagging of images captured by smartphone's camera app. & 3.23 & 1.48 \\
 & PA7 & I check which apps are running in the background. & 3.33 & 1.14 \\
 & PA8 & I check my smartphone's privacy settings. & 3.31 & 1.17 \\
 & PA9 & When receiving a link from an unknown source via SMS, I click the link immediately. (r) & 1.67 & 1.04 \\
\hline
\multirow{2}{*}{Update} & UP1 & When I'm prompted about a software update on my smartphone, I install it right away. & 3.3 & 1.13 \\
 & UP2 & I make sure that the smartphone applications I use are up-to-date. & 3.61 & 0.92 \\
\end{tabular}
}%
\end{table*}

To answer these research questions, we adopted a psychometric approach. Psychometrics is a scientific approach of quantifying human psychological attributes, such as personality traits, cognitive abilities, and social attitudes~\cite{michell2008psychometrics}. A well-developed and widely-used security-related psychometric measurement is the ``Security Behavior Intention Scale'' (SeBIS) developed by Egelman and Peer~\cite{egelman2015scaling}. They conceptualized users' \textit{general} security behavior as a psychological construct instead of an actual behavior. We follow the same approach to conceptualize users' \textit{smartphone} security behavior as a psychological construct of behavioral intent that can be predictive of actual behaviors. 

We develop the Smartphone Security Behavioral Scale (SSBS), a new measurement for assessing users' behavior intentions to comply with smartphone security practices. When developing a new scale, it is important to evaluate three psychometric properties of the measurement: dimensionality, scale reliability, and convergent validity~\cite{netemeyer2003scaling}.

\textit{Dimensionality.} Identifying dimensionality of a construct is a critical part of scale development because whether the construct is uni-dimensional or multi-dimensional will affect the structure and computing approach of scale~\cite{netemeyer2003scaling}. There are two statistical approaches to determine dimensionality based on the use case. If the goal of testing is to ‘explore' the unknown dimensions of a construct, the Exploratory Factor Analysis (EFA) is an appropriate method to use. Then, if the goal is to `confirm' or examine the existing dimensions of a construct, the Confirmatory Factor Analysis (CFA) is a standardized way to test the fitness of the model.  

\textit{Scale Reliability.} In psychometrics, reliability represents the consistency of a measurement, which can be evaluated in various ways~\cite{netemeyer2003scaling}. In this study, we focus on assessing ``internal consistency'' of the scale to determine if multiple items in a scale measure the same construct by examining Cronbach's alpha~\cite{cronbach1951coefficient}. Cronbach's alpha is the mean of all possible coefficients among items~\cite{cortina1993coefficient}. The cut-off point of Cronbach's alpha is 0.70~\cite{nunnally1978} that refers to the acceptable internal consistency of the scale. In addition, considering the numbers of items can affect the score of Cronbach's alpha~\cite{cortina1993coefficient, tavakol2011making}, we also report the mean of inter-item correlation (ITC), which is the average pairwise correlation among all items and provides a direct indicator of homogeneity~\cite{cohen1996psychological}. 

\textit{Construct Validity.} This refers to the degree to which a measurement truly reflects the concept being examined~\cite{calder1982concept}. One approach is to evaluate convergent validity between the developing scale and an existing scale which measures the same construct as the developing scale~\cite{nunnally1994psychometric, netemeyer2003scaling}. Convergent validity is measured by correlational coefficients between the new measure and an existing measure. In our study, we evaluated the convergent validity between our scale and SeBIS~\cite{egelman2015scaling} and tested if our scale measures similar constructs of security behavior. 

Since there has been a well-established computer security behavioral intentions scale (SeBIS), our first step was to examine if the dimensionality of SeBIS can be applied to users' smartphone security behavior intentions. Our findings indicate different dimensions of smartphone security. Therefore, we followed a standardized procedure of scale development proposed by Netemeyer~\cite{netemeyer2003scaling}. Our procedure of scale development is summarized as follows:
\begin{enumerate}
\item Testing the fitness of dimensional model of SeBIS on smartphone security behavior by applying CFA.
\item Defining the construct that the scale attempted to measure and generating a list of candidate questions.
\item Extracting the dimensional components of the scale by performing EFA and reducing the set of items.
\item Finalizing the scale by conducting CFA to confirm the fitness of the new scale to the intended factorial model. 
\end{enumerate}


\vspace{5pt} \noindent \textbf{Methodology.} The goal of this study is to develop a measurement to assess users' security behavior intentions to comply with smartphone security advice recommended by security professionals. We conducted a two-phase online survey study. In phase-1, we tested the four dimensions used in SeBIS~\cite{egelman2015scaling}. Our results suggest the possibility of improving on the four dimensions of SeBIS when specializing to smartphone security behavior intentions. That is, users' smartphone security behavior intentions could be different from the general computer security behavior intentions. We therefore conducted a phase-2 study to develop a new measurement for smartphone security behavior intentions. 

For both phases, we recruited participants from the United States via Amazon Mechanical Turk. To ensure the quality of data, we integrated attention-check questions in each section of the survey. The attention-check questions were randomly inserted in the questionnaire and had similar format to other questions. Participants were required to select the choice required in the statement (for instance, \textit{I go to grocery shopping on every Thursday. Please select `Never'}). We removed the responses from participants who failed to correctly answer attention-check questions. We next describe the details of study design and results for each phase of the study.

%% file: phase_1_study.tex
\subsection{Survey design and item generation}
We first developed a measurement, which we call \textit{smartphone-SeBIS}, based on the four dimensions of SeBIS: device securement, password management, proactive awareness, and update. We generated items by revising each question in SeBIS to adapt to a smartphone context. For example, we changed the wording of questions from `computer' to `smartphone'. However, we encountered two challenges when using this approach. First, we found that certain questions could not be readily applied to smartphones. Second, certain common smartphone-specific security features were not included in SeBIS, such as biometrics, usage of applications, and app permissions. To capture a more comprehensive view of users' smartphone security behaviors, we recruited security experts who independently went over each item of the first version of smartphone-SeBIS and considered how to revise old items and add new items to the survey. Overall, we had four types of item modifications: word/phrase substitution, word/phrase revision, item deletion, and item addition. 

\textit{Word/Phrase Substitution:} we substituted words indicating the context of a laptop or desktop machine to specifically describe a smartphone. For instance, to capture the same behavior on a smartphone device, we substituted the word ``smartphone" for ``laptop or tablet'' in the item ``\textit{I use a password/passcode to unlock my laptop or tablet.}''

\textit{Word/Phrase Revision:} some items could not be made smartphone-specific with simple substitutions. For example, ``\textit{I do not change my passwords, unless I have to}''. This was revised to the following: ``\textit{I regularly change my password for online services/accounts using my smartphone}'', where we specified the password target to avoid confusion, and turn the negative statement into a positive statement. We did this since participants might be biased toward taking a defensive stance against the negative behavior. 

\textit{Item Deletion:} Some of SeBIS items are not applicable to the smartphone context. For instance, the item ``\textit{When browsing websites, I mouseover links to see where they go, before clicking them}'' is not applicable on mobile devices since the pointing mechanism on smartphones is different (mouse or trackpad for desktops/laptops vs finger or stylus on mobile devices). Such items were removed from the survey.

\textit{Item Addition:} Several important smartphone security behaviors were not specified or included in SeBIS. For instance, significant security mechanisms introduced by Original Equipment Manufacturers (OEMs), or by the research community, become obsolete if the user roots (or jailbreaks) their smartphone. This is an important ``device securement'' measurement to take. Moreover, on smartphones, user privacy is preserved through a permission system that allows users to determine what device and personal information each installed third-party app can access. This mechanism can also be compromised if users become inattentive to permission requests or if they never revoke permissions from apps~\cite{wijesekera2015android, felt2012android}. To address such phenomena, we added relevant smartphone-specific items into the survey. 


Smartphone-SeBIS resulted from this exercise. It has a total of 20 items targeting smartphone security behaviors (see Table~\ref{table2}). We administered smartphone-SeBIS through an online survey (Amazon MTurk) where participants were asked to answer on a 5-point Likert-type scale (from `Never' to `Always'). To avoid the priming effect of social desirability, we advertised our study as `the use of smartphone and mental health wellness'. In the survey questionnaire, we included three measurements of mental health wellness (Patient Health Questionnaire (PHQ-9), Insomnia Severity Index (ISI), and Internet Addiction Test (IAT)), one measurement of general computer security (SeBIS), and our preliminary measurement  smartphone-SeBIS. To prevent the order effect, all five survey sections were randomized. After answering these five sections, participants were asked about their demographics.

\textit{Survey demographics:} we recruited a total of 100 participants. Ages of participants were between 18 to 71 ($\mu$=36.2, $\sigma$=11.4), and 41 of them are female (41\%). Thirteen percent of our participants had a high school diploma (n=13); 36\% had some college or associate degree (n=36); 36\% had bachelor's degree (n=36); and 15\% had a graduate or professional degree (n=15). The average time to take the survey was 11.7 minutes. Each participant was rewarded \$0.75 for their participation.

\subsection{Results}
The analysis shows that the internal reliability of the 20-item smartphone-SeBIS is below the recommended cutoff point by Nunnally (1978) (Cronbach's $\alpha$=.67<.70)~\cite{nunnally1978}. We further conducted Confirmatory Factor Analysis (CFA) to examine whether our measurement of the construct is consistent with SeBIS by the goodness-of-fit of data to the latent variable model. We used several tests to determine the goodness-of-fit of data to the model of SeBIS, including the Comparative Fit Index (CFI), the Tucker-Lewis Index (TLI), the Root Mean Square Error of Approximation (RMSEA), and the Standardized Root Mean Square Residual (SRMR). 
According to our results, the CFI and TLI were 0.565 and 0.490, which are below the cutoff (0.90) recommended by Netemeyer et al.~\cite{netemeyer2003scaling}. Furthermore, our RMSEA and SRMR are 0.127 and 0.152 respectively, which are above the recommended cutoff points (a cutoff of 0.06 for RMSEA and 0.08 for SRMR~\cite{hu1999cutoff}). These results indicate poor goodness-of-fit of our data to smartphone-SeBIS. That is: \textit{the revised four dimensions of smartphone-SeBIS might not be the best fit for assessing users' smartphone security behavior intentions.}

%% file: phase_2_study.tex
\subsection{Survey design and item generation}
To develop a new scale to measure users' smartphone security behavior intentions we employed the approach used by Egelman and Peer (2015). We first generated a list of smartphone security behaviors and collected data on Amazon MTurk. Then we conducted an Exploratory Factor Analysis (EFA) to extract the effective items for assessing users' smartphone security behavior. 

\textit{Item generation:} According to Egelman and Peer~\cite{egelman2015scaling}, the metric of security behaviors should be ``applicable'' to and ``widely accepted'' by the majority of users. Therefore, we generated the list of smartphone security behaviors based on the views of security professionals. We invited 35 students who majored in computer science and received substantial professional training in cybersecurity to list 10 of the most important smartphone security behaviors. Then two security researchers categorized these and generated a list of behaviors. The researchers also examined public security advice for smartphone security by the United States Computer Emergency Readiness Team (US-CERT) to ensure no important behaviors are missing from the list. Then, five security experts went through the list to determine if any item violated the principles of applicability and acceptance. Our initial list contained 45 behaviors. We then translated these behaviors into personal statements. Participants were asked to read and rate each statement on a five point scale of frequency (From `Never' to `Always').

\textit{Survey Demographics:} we collected 487 responses via Amazon MTurk. This is larger sample than the sample size recommended by Hair \textit{et al.} (minimum of 5 participants per item)~\cite{hair2013multivariate}. The average age of participants was 34.6 and 44.8\% are female (41\%). About 11\% of our participants had high school diplomas (n=54); 29\% had some college or associate degree (n=142); 49.5\% had a bachelor's degree (n=241); and 10.3\% had a graduate or professional degree (n=50). The average time taken to complete the surveay was 6.3 minutes. Participants were paid \$0.75 after completing the survey and passing the validation checks.

\subsection{Results}
\subsubsection{Exploratory Factor Analysis}
Our analysis of Kaiser-Meyer-Olkin (KMO) test is 0.92 indicating the high sampling adequacy of variables, which suggest suitability for further factor analysis.  Considering a large set of items, our approach was to refine our scales until the loading of each item was above 0.5 and was twice more than its loading on other components~\cite{saucier1994mini} after a Varimax rotation. Furthermore, we used optimal coordinate to determine the optimal number of factors, which is a non-graphical approach for factor determination~\cite{raiche2013non}. The optimal coordinate is a determined point where the predicted eigenvalue is not greater than or equal to the mean eigenvalue by performing linear regression analysis of the last and (i+1)th eigenvalue~\cite{raiche2013non}. By using optimal coordinates, we can overcome a limitation of subjective and unclear decision-making about the number of components to retain~\cite{raiche2013non}.  

We performed three rounds of EFA to finalize our scale of smartphone security behavioral intention. In our first round of EFA, we first conducted Principal Component Analysis (PCA) and extracted five components in EFA based on optimal coordinate analysis. Then we excluded 27 items based on aforementioned loading criteria. In the second round of EFA, we followed the same procedure and extracted three components in EFA. Then, 3 items were excluded from the list of items. In the third round of EFA, we also followed the same procedure that performed in the last two rounds and extracted 2 components in EFA and retained the remaining 14 items. The final set of items and their rotated factor loadings are displayed in Table~\ref{table4}. According to each component's items, two themes appeared: \textit{technical} (\textit{e.g.}, using VPN and anti-virus app) and \textit{social} (\textit{e.g.}, verifying the source of texts before sharing, deleting suspicious communication) approaches. Our scale includes these two subscales.

\subsubsection{Reliability of the Scale}
We adopted the same approach used by Egleman and Peer~\cite{egelman2015scaling} to examine the reliability of the scale based on three metrics. First we employ Cronbach's $\alpha$, which is commonly used to assess internal consistency of a group of items. As shown in Table~\ref{table4}, the Cronbach's $\alpha$ for the full scale is 0.80. For subscales of technical and social approaches are 0.84 and 0.79 respectively. Our scale met the criteria of internal consistency that requires both full scale and all subscales to be above 0.7~\cite{nunnally1994psychometric, mckinley1997reliability}. Then we leverage item-total correlation (ITC), which is the Pearson correlation between each item and the mean of all other items. All of our items' ITC are above the recommended threshold of 0.2~\cite{everitt2010cambridge}. 

While assessing the reliability of the scale, it is also important to examine the diversity of the items of a scale and prevent the redundancy of the items~\cite{allen2001introduction}\ignore{Allen \& Yen, 2002}. Toward this, we computed the average inter-item correlation (IIC) that not only evaluates the internal consistency but also tests the degree of redundancy of a set of items on a scale~\cite{cohen2005psychological, piedmont2014inter}. The recommended correlational coefficient of IIC is between 0.20 and 0.40, which suggests that the items contain sufficient diversity of variance while they are still representative of the same construct~\cite{piedmont2014inter}. ITC of both subscales fall within the range, which indicates the adequate level between consistency and diversity. Based on these three metrics, our full scale and sub-scales exhibit high reliability. 

\subsubsection{Convergent Validity: Correlation with SeBIS }
To ensure that we assess the construct of users' security behavior, we measured the convergent validity of our scale and SeBIS. Convergent validity is a type of criterion validity that evaluate if a developed scale measures the same construct of the `criterion' scale. We used SeBIS as our criterion because it is the only measure with high reliability for assessing security behavioral intentions. We collected a new dataset with 66 participants who completed both SeBIS and Smartphone Security Behavior Scale (SSBS). Then we conducted Pearson's correlation between SeBIS and SSBS. The average score of SeBIS has significantly positive correlation with the average score of SSBS (r=.403, p=.0008). In addition, results show the positive significant correlation between the subscales of SeBIS and SSBS (see Table~\ref{table5}). These findings suggest that \textit{participants who showed higher intentions in protecting their general security were also more likely to protect their smartphone security}. This confirms that our scale is measuring a similar construct with SeBIS, that of security behavior intentions. 

\begin{table}[]
\centering
\caption{Pearson's Correlation between SeBIS and SSBS}
\label{table5}
\resizebox{\columnwidth}{!}{%
\begin{tabular}{lll}
 & \multicolumn{2}{l}{Correlation coefficient (p-value)} \\
 \hline
SeBIS           /         SSBS & Technical approach & Social approach \\
\hline
Device securement & -.017 (p=.896) & .060 (p=.628) \\
Password generation & .290 (p=.018) & .229 (p=.064) \\
Proactive awareness & -.090 (p=.471) & .614 (p\textless{}.0001) \\
Update & .301 (p=.014) & .431 (p=.0003)\\
\end{tabular}%
}
\vspace{-10pt}
\end{table}

\subsubsection{Confirmatory data analysis}
Our final step is to examine the goodness of fit of SSBS with the hypothesized latent components by performing Confirmatory Factor Analysis (CFA). We collected a new dataset with 358 U.S. participants from Amazon MTurk in this final round of survey. To avoid the priming effect on participants' responses, we adopted the same approach employed in the phase-1 study to advertise the survey as a research related to users' mobile phone usage and mental wellbeing. We then included three measurements of mental health wellness (PHQ-9, Insomnia Severity Index, and Internet Addiction Test), SeBIS, SSBS and demographics items in the survey. All survey sections were randomized to avoid the order effect on participants' responses. Moreover, we include attention-check questions in each survey section and removed the responses which failed to answer them correctly. In terms of demographics, 38\% (n=136) of our participants are female and the average age of participants was 35.3 ($\sigma$=10.6). Each participant was paid \$1 after completing the survey. 

The reliability of full SSBS is 0.79 and it is 0.81 for the \emph{Technical} subscale and 0.85 for the \emph{Social} subscale. We conducted PCA with a Varimax rotation and extracted two components. The results show that all items were loaded on the same unique component as found in the previous EFA. Then we conducted CFA to examine the goodness-of-fit of the two-component model for users' smartphone security behavior intentions. We used the same approach employed in Phase-1 study, we performed multiple test to determine the goodness-of-fit of our data to the model, including Comparative Fit Index (CFI), the Tucker-Lewis Index (TLI), the Root Mean Square Error of Approximation (RMSEA), and the Standardized Root Mean Square Residual (SRMR). Based on the analysis, the CFI and TLI were 0.954 and 0.942, which are above the cutoff (0.90) recommended by Netemeyer et al.~\cite{netemeyer2003scaling}. Also, the RMSEA and SRMR is 0.054 and 0.059 respectively. Both scores are below the cutoff points recommended by~\cite{hu1999cutoff}. Our results show a well goodness-of-fit of our data to our hypothesized two-component model. We also performed Pearson's correlation between the two subscales and found no significant correlations.

\begin{table*}[]
\small
\centering
\caption{Factor loadings and reliability statistics of finalized scale}
\label{table4}
\resizebox{\textwidth}{!}{%
\begin{tabular}{l|l|l|l|l}

\textbf{ID}       & \textbf{Item}                                                                                                                                      & \textbf{Technical} & \textbf{Social} & \textbf{Inter-total correlation} \\ \hline
T1               & I reset my Advertising ID on my smartphone.                                                                                                        & .787               &                 & 0.52                             \\ 
T2               & I hide device in my smartphone’s bluetooth settings.                                                                                              & .639               &                 & 0.47                             \\ 
T3               & I change my passcode/PIN for my smartphone’s screen lock at a regular basis.                                                                      & .629               &                 & 0.51                             \\ 
T4               & I manually cover my smartphone’s screen when using it in the public area (e.g., bus or subway).                                                    & .621               &                 & 0.55                             \\ 
T5               & I use an adblocker on my smartphone.                                                                                                               & .614               &                 & 0.51                             \\ 
T6               & I use an anti-virus app.                                                                                                                           & .612               &                 & 0.53                             \\ 
T7                & I use a Virtual Private Network (VPN) app while connected to a public network.                                                                     & .604               &                 & 0.42                             \\ 
T8                & I turn off WiFi on my smartphone when not actively using it.                                                                      & .544               &                 & 0.47                             \\ 
S1               & I care about the source of the app when performing financial and/or shopping tasks on that app.                                                         &                    & .723            & 0.24                             \\ 
S2                & When downloading an app, I check that the app is from the official/expected source.                                                                &                    & .677            & 0.36                             \\ 
S3                & Before downloading a smartphone app I ensure the download is from official application stores. &                    & .677            & 0.21                             \\ 
S4               & I verify the recipient/sender before sharing text messages or other information using smartphone apps.                                             &                    & .609            & 0.41                             \\ 
S5               & I delete any online communications (i.e., texts, emails, social media posts) that look suspicious.                                                 &                    & .552            & 0.25                             \\ 
S6               & I pay attention to the pop-ups on my smartphone when connecting it to another device (e.g. laptop, desktop).                                         &                    & .526            & 0.39                             \\ \hline
\multirow{2}{*}{} & \textbf{Cronbach's alpha}                                                                                                                          & 0.84               & 0.79            &                                  \\ \cline{2-5} 
                  & \textbf{Inter-item correlation}                                                                                                                    & 0.40               & 0.39            &                                  \\ 
\end{tabular}
}
\vspace{-5pt}
\end{table*}



%% file: mhi_method.tex
SSBS can be an enabler of studies in understanding the repercussions of smartphone security behavior intentions. As an example, we choose to study the relationship between  smartphone security behavior intentions and mental health issues (MHIs). We chose this application for the following reasons.

(1) MHIs are prevalent. Mental health issues (MHIs) affect millions of people around the globe. According to NAMI, 1 in 5 adults in the US experiences mental illness in a given year, and 1 in 5 youth aged 13-28 experiences a severe mental disorder art some point in their lives. Also, 16 million adults in the US had at least one major depressive episode in 2018.

(2) Given existing evidence, we hypothesize that MHIs correlate with security behavior. Certain personality traits appear more likely to lead to future mental illness~\cite{personalityMHI}. Given that people afflicted with MHIs tend to experience distinct psychological constructs, and because some personality traits (with their own distinct psychological constructs) are shown to cause security risks~\cite{uffen2012personality}, \cite{jeske2016exploring}, this led us to hypothesize that MHIs could also correlate with security behavior.

(3) We are explicitly interested in better understanding the relationship of MHIs with \textit{smartphone} security behavior intentions since this can have important socioeconomic applications. For example, if people exhibiting a mental health issue tend to follow bad security hygiene on their smartphones in a bring-your-own-device (BYOD) enterprise setting, then enterprises can use this to better estimate risk of security breaches and also focus their security educational programs which will in turn save the enterprise millions of dollars---according to the Anti-Phishing Working Group direct and indirect phishing incidents led to data breaches costing enterprises more than \$3.86 million dollars. On the other hand, if security behaviors are indicative of a person's mental health this could be weaponized by advertisers or insurance companies aiming to profile and target vulnerable groups of the population~\cite{demetriou2016free}. Understanding the latter will allow us to better protect consumers on their personal mobile devices from such aggressive third-parties. Next we present our studies which apply SSBS to tackle three concrete research questions.

\noindent$\bullet$\textbf{ RQ3:} Do people with MHIs have distinct smartphone security behavior?

\noindent$\bullet$\textbf{ RQ4:} Can mental health issues be potentially leveraged as an important predictor of smartphone security behaviors?

\noindent$\bullet$\textbf{ RQ5:} Can smartphone security attitudes be potentially leveraged as an important predictor of mental health issues?

\ignore{
\begin{itemize}
    \item \textbf{RQ3:} Are there significant differences between users with and without mental health issues on their smartphone security behaviors?
    \item \textbf{RQ4:} Can mental health issues be potentially leveraged as an important predictor of smartphone security behaviors?
    \item \textbf{RQ5:} Can smartphone security attitudes be potentially leveraged as an important predictor of mental health issues?
\end{itemize}
}

To address RQ3, our survey results were analyzed using statistical comparison test to examine the relationship between prevalent MHIs and smartphone security attitudes. Participants' responses were categorized into groups showing proclivity for a particular mental condition. Each group's smartphone security behavioral intentions (from SSBS) were compared with a control group, which does not show the same proclivity. To measure these proclivities we carefully select instruments which can measure MHIs which are prevalent globally. This would render our results more relevant to current societal issues and facilitate recruitment. We hence focused on depression which affects 300 million people worldwide~\cite{whodepression} and insomnia which affects half the global population~\cite{bhaskar2016prevalence}. Lastly, we also include a form of addiction, Internet Addiction, which gradually become prevalent in our society~\cite{kuss2016internet, shao2018internet}. 


For depression we used the Patient Health Questionnaire---PHQ-9~\cite{kroenke2002phq, lowe2004measuring}. It's brevity, construct and criterion validity render it an ideal candidate for efficient distribution~\cite{kroenke2002phq}. Furthermore,  previous work has confirmed that PHQ-9 is suitable to be administered to participants~\cite{lowe2004measuring}. The insomnia severity index (ISI)~\cite{bastien2001validation, morin2011insomnia} was also found to be a reliable self-report measure for evaluating sleep difficulties. For internet addiction, we adopted the \textit{Internet Addiction Test} (IAT)~\cite{widyanto2004psychometric}.




To address RQ4, we use a generalized linear model (GLM) to assess the predictive influence of users' mental health issues on their smartphone security behavioral intentions. GLM is a flexible generalization of ordinary linear regression model that allows response variables to have non-normal distribution. Considering the continuous nature of our response variable, we predicted the response variable via a link function, which specifies the link between random and systematic components. The link  shows how the linear predictor of explanatory variables relates to the expected value of the response.

To address RQ5, we utilize SSBS to construct features to train and evaluate off-the-shelve machine learning binary classifiers. Each classifier is tasked with predicting whether an individual is likely to exhibit a specific mental health issue or not.  To better explain the classifiers' performance and understand which security attitudes contribute the most to the prediction task, we further evaluated each individual feature's influence in the prediction performance. 

%% file: correlations.tex
To study the relation between smartphone security behavioral intentions and MHIs, we first convert the MHI scores (PHQ9, ISI, IAT) into binary values. For each MHI, we decide a threshold such that any score equal or above it would be converted to a value of 1 (equivalent to having moderate or severe symptoms) and any value below it would be converted to a 0 (equivalent to not having moderate or severe symptoms). For Depression the threshold was set to a PHQ9 score of 10~\cite{levis2019accuracy}, for Insomnia it was set to an ISI score of 8, and for Internet Addiction to an IAT score of 50. Moreover, since \mSebisName{} has distinct items which are factored together to measure a specific cybersecurity tendency, we average the Likert scores of the items for each of its two subscales. This allows us to derive for each participant an average score for each of the two distinct underlying constructs of \mSebisName{}, namely \emph{Technical} and \emph{Social}. In total we observed 190/358 (53.1\%) with high IAT score, 117/358 (32.7\%) participants with high PHQ9 score and 197/358 (55.0\%) with high ISI score. 

Considering the non-normal distribution of our smartphone security behavioral scales, we conducted non-parametric test to examine the difference in mental health issues and smartphone security behaviors by using the Wilcoxon Rank Sum test. Below we elaborate on our analysis results.



\textbf{Depression and Smartphone Security Behavior.}
Our results show that participants with depression indicators reported significantly higher behavioral intentions on adopting a technical strategy to protect their smartphones than those with low depression scores (W=14524, p<.0001). Conversely, participants with high depression scores reported significantly lower social security behavioral intentions than those with low depression scores (W=9022.5, p=.019).

\textbf{Insomnia and Smartphone Security Behavior.}
Similarly, we found that participants with high ISI score reported significantly higher behavioral intentions on adopting a technical strategy to protect their smartphones than those with low ISI score (W=19517, p=.0002). We also found that participants with high ISI score reported significantly lower behavioral intentions on using social strategies to protect their smartphones than those with low insomnia scores (W=12118, p=.0001).

\textbf{Internet Addiction and Smartphone Security Behavior.}
For internet addiction, our results reveal a similar trend. Participants with high IAT score reported significantly higher behavioral intentions on adopting technical strategies to protect their smartphones than those with low internet addiction scores (W=19684, p=.0002). On the other hand, participants with high IAT score reported significantly lower behavioral intentions in adopting social strategies than those with low IAT score (W=9049.5, p<.0001).




%% file: predicting_sec.tex
To understand whether mental health issues could potentially predict smartphone security behavior attitudes (RQ4), we conducted Generalized Linear Modeling (GLM) to examine the predictive ability of the MHIs (depression, insomnia, and internet addiction) on users' mobile security behavioral intentions by controlling their age, gender, and smartphone usage time. 

The first model predicts users' behavioral intentions toward the \emph{Technical} approach of smartphone security. Our results show that \textit{participants with high PHQ9 score are significantly more likely to show higher behavioral intentions toward technical strategies for smartphone security than those with low PHQ9 score ($\beta$=0.361, t-value=2.79, p=.006)}. Yet, there is no significant predictive power of insomnia and internet addiction on technical smartphone security behaviors.

Focusing on the \emph{Social} aspect we found that \textit{participants with low IAT score are significantly more likely to show higher behavioral intentions toward using social strategies to protect their smartphones than those who have high IAT score ($\beta$=0.453, t-value=4.55, p<.0001)}. However, there are no statistically significant predictive differences in depression and insomnia w.r.t social behaviors.

These results reveal an opportunity for enterprises to focus their expensive security educational programs. For example, educating employees with high PHQ9 scores about taking technical measures to protect their smartphones can be highly effective. Analogously, the same might be true when employees with high IAT scores are subjected to similar programs for training on social smartphone security strategies. On the other hand, adversaries might also leverage these results when targeting users through technical means (malicious mobile apps) or social engineering attacks (spam and phishing attacks).

%% file: predicting_mhi.tex

Adversaries, advertising or insurance companies might also try to identify or profile vulnerable groups. In this section, we present how we model the prediction of MHIs based on security behavior intentions and investigate whether such models can be leveraged to launch profiling or targeted attacks.

\subsection{Methodology}
We predict an MHI issue (Depression, Insomnia, Internet Addiction) based on the smartphone security behavioral scale (\emph{SSBS}). This resulted in a total of three classification tasks. Each classification task was tackled using seven classification algorithms (\textit{Logistic Regression}, \textit{SVM}, \textit{Random Forest}, \textit{KNN}, \textit{Stratified}, \textit{Most Frequent}, \textit{Random}). The first four are common classification algorithms that are well suited for our problem. 


We then implemented three simpler models which we use as baselines and for helping interpretations of the results of the above classifiers. The stratified model, generates predictions by respecting the training set's class distribution, the most frequent model always chooses the label to be the most frequently used label in the training set, and lastly the random model assigns prediction uniformly at random.



Due to our relevantly small dataset we convert our tasks into binary classification problems. Thus, for each MHI issue, we decide a threshold such that any score equal or above it would be converted to a value of 1 (equivalent to having moderate or severe symptoms) and any value below it would be converted to a 0 (equivalent to not having moderate or severe symptoms). \ignore{For Depression the threshold was set to a PHQ9 score of 10, for Insomnia it was set to an ISI score of 8, and for Internet Addiction to an IAT score of 50.} In terms of the independent variables (features) we use the participant responses in the individual items of SSBS. 

Then, split our participant responses into 75\% (n=268) to be used for training the classification models and 25\% (n=90) for testing them. Table~\ref{tab:eval_data_distr} lists the label count in the training and testing data for each MHI. We observe the low number of participants with positive depression labels (label=1). This is known as the class imbalance problem and in our case it would prevent the depression classifiers from being able to predict moderate or severe depression symptoms. To tackle imbalanced datasets, one can either undersample the majority class (label=0 in our case) or oversample the minority class (label=1). We eliminated the undersampling approach as this would further reduce our data. In terms of oversampling the minority class, one could trivially replicate the minority class multiple times\ignore{ until we have a balanced number of 0 and 1 labels}. However this approach could suffer from overfitting to a few minority class cases.\ignore{ we can observe}. A better approach (SMOTE) randomly picks a datapoint in the minority class and adds synthetic points placed between it and its k-nearest neighbors~\cite{chawla2002smote}. We used SMOTE on our training data which resulted in an equal number of label values (179). To determine the models' hyperparameters we run 10-fold cross validation for all models for each classification task. 


\begin{table}[]
\small
\center
\caption{Data Distributions}
\label{tab:eval_data_distr}
\resizebox{\linewidth}{!}{%
\begin{tabular}{l|l|l|l|l}

            & \multicolumn{2}{c|}{\textbf{Training Data (n=268)}}               & \multicolumn{2}{c}{\textbf{Testing Data (n=90)}}                 \\ \hline
            & \multicolumn{1}{c|}{\textbf{0}} & \multicolumn{1}{c|}{\textbf{1}} & \multicolumn{1}{c|}{\textbf{0}} & \multicolumn{1}{c}{\textbf{1}} \\ \hline
\textbf{Internet Addiction} & 124 (46\%)                      & 144 (54\%)                      & 44 (48.9\%)                     & 46 (51.1\%)                     \\ \hline
\textbf{Depression}  & 179 (67\%)                      & 89 (33\%)                       & 62 (68.9\%)                     & 28 (31.1\%)                     \\ \hline
\textbf{Insomnia}  & 125 (47\%)                      & 143 (53\%)                      & 36 (40\%)                       & 54 (60\%)                       \\ 
\end{tabular}
}
\vspace{-10pt}
\end{table}

\subsection{Results}
Table~\ref{tab:eval_class_perf} summarizes our results. We evaluate all models on the testing dataset (unseen data) and report each classifier's AUC, precision and recall. Intuitively AUC is a measure of how well the model can distinguish between our two classes---the higher the AUC the better the model is at predicting participants with high MHI scores. Precision tells us how many of the participants we labelled as positives are actually positives, whereas recall tells us how many of all the positive participants we managed to correctly label as positives.


We observe that the most straightforward models (\textit{Stratified}, \textit{Most Frequent}, \textit{Random}) have very low AUC which means they are bad in distinguishing the positive (1) from the negative labels (0). Our more sophisticated models trained on SSBS items always improve the quality of the predictions. Due to space limitations we omit their detailed results from Table~\ref{tab:eval_class_perf}.


\textbf{Internet Addiction:} We observe that SVM, Random Forest and KNN achieve very good separability (AUC=70.3\%, 70.9\%, 72.1\% respectively). However, SVM and Random Forest tend to miss a lot of participants with high scores (recall=59\%). On the other hand KNN achieves both good recall(72\%) and good precision (70\%).  

\textbf{Depression:} SVM, Random Forest and KNN show good label distinguishability (AUC=75.1\%, 74.5\%, 75.8\% respectively) and KNN being the only one with high recall (71\%). In terms of precision Random Forest is significantly better than the rest (64\%). This shows that KNN might be preferable when one wants to prioritize identifying most of the positive cases than making sure that the positive labels are correctly assigned. Advertisers might choose this overestimation strategy when launching campaigns where coverage is important. On the other hand, insurance companies  might focus more on an approach that does not unjustifiably overcharge clients and thus might choose the Random Forest approach.

\textbf{Insomnia:} All classifiers struggle to achieve good separability (only Random Forest comes close to 70\%) and exhibit low recall (high number of false negatives). However, Logistic Regression, SVM and Random Forest achieve high precision which indicates they could be potentially leveraged by insurance companies or even by advertisers in targeted campaigns.

\begin{table}[]
\center
\caption{Predicting MHI scores from SSBS features (\%)}
\label{tab:eval_class_perf}
\begin{tabular}{l | l | l | l }
\textbf{Internet Addiction} &  \textbf{Precision} & \textbf{Recall}  & \textbf{AUC}  \\ \hline
Logistic Regression         &         68              &      50           &           66.2        \\
SVM                         &             66              &      59           &            70.3        \\
Random Forest               &           64              &      59           &            70.9        \\
KNN                         &               70            &        72         &                72.1      \\ \hline
\textbf{Depression} &  \textbf{Precision} & \textbf{Recall} &  \textbf{AUC}  \\ \hline
Logistic Regression         &               52            &        57         &            68.3       \\
SVM                         &             54              &      50           &        75.1         \\
Random Forest               &             64              &      57           &         74.5         \\
KNN                         &             54              &      71           &         75.8         \\ \hline
\textbf{Insomnia} &  \textbf{Precision} & \textbf{Recall} &  \textbf{AUC}  \\ \hline
Logistic Regression         &             78              &      54          &          62.3        \\
SVM                         &            71              &      46           &         62.8         \\
Random Forest               &            73              &      56           &         66.2         \\
KNN                         &             67              &      56           &         62.1         \\
\end{tabular}
\vspace{-10pt}
\end{table}

\subsection{Analyzing the Influence of Individual Items}
\begin{figure*}[!htb]
\minipage{0.33\textwidth}%
  \includegraphics[width=\linewidth]{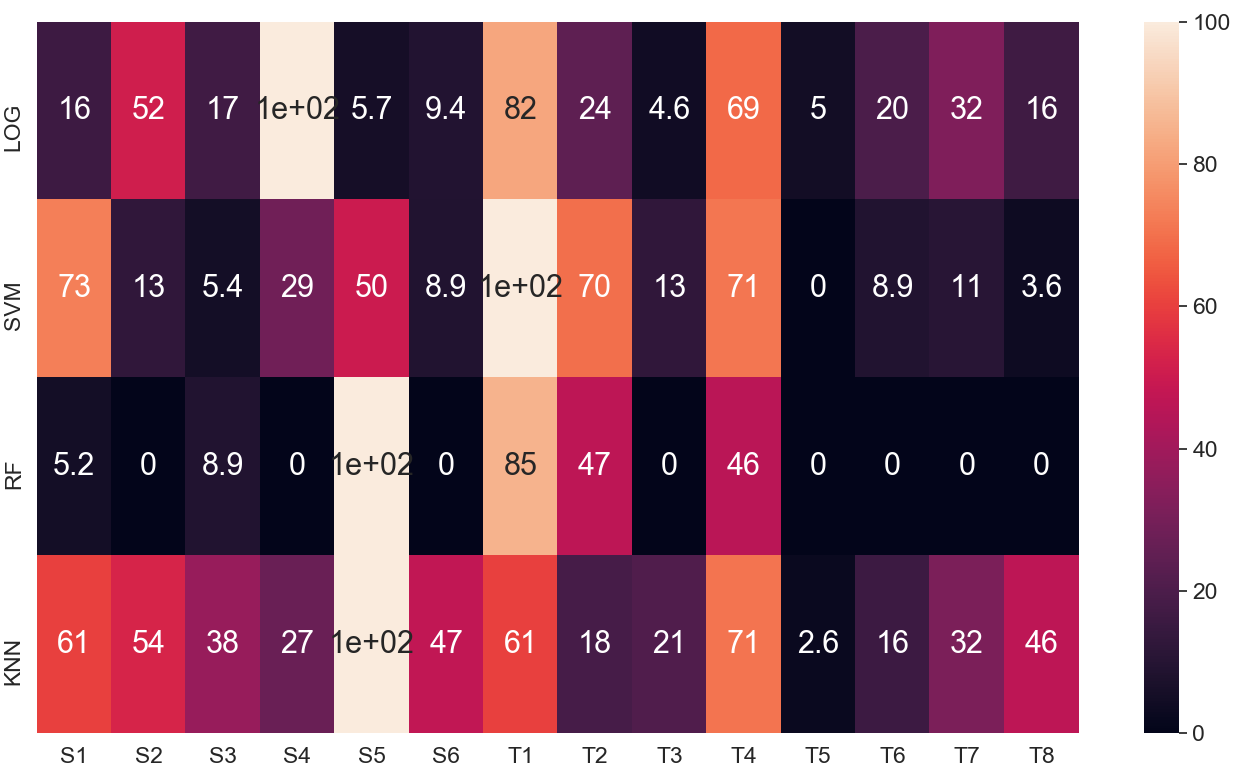}
  \subcaption{Internet Addiction}\label{fig:fiIaMsebis}
\endminipage \hfill
\minipage{0.33\textwidth}
  \includegraphics[width=\linewidth]{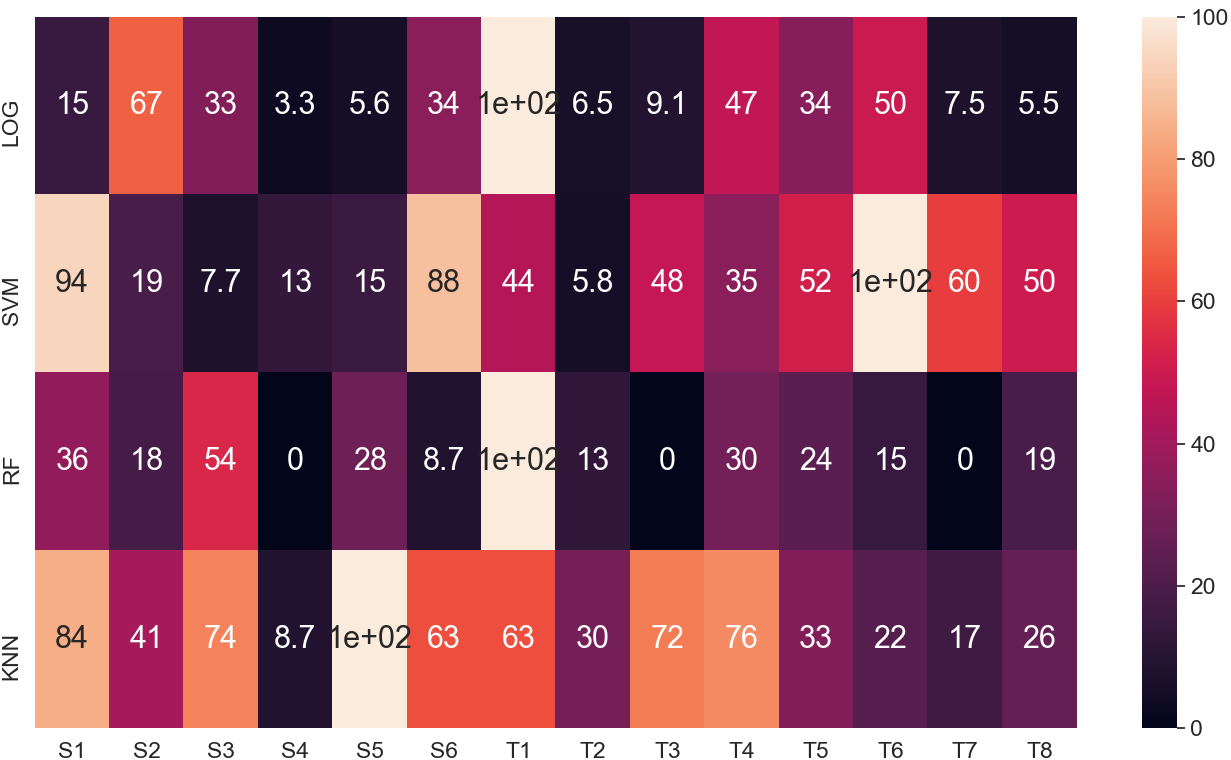}
  \subcaption{Depression}\label{fig:fiPhq9Msebis}
\endminipage\hfill
\minipage{0.33\textwidth}
  \includegraphics[width=\linewidth]{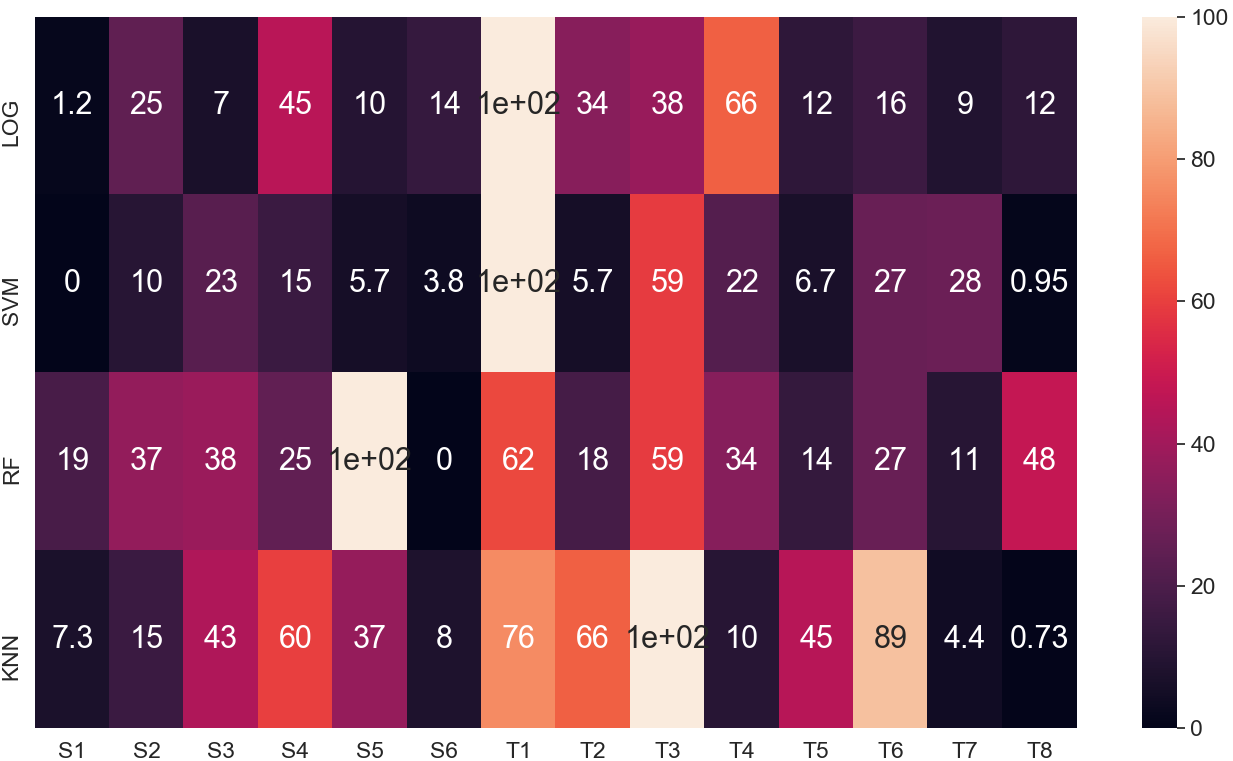}
  \subcaption{Insomnia}\label{fig:fiInsomniaMsebis}
\endminipage
\caption{Relative Feature Importances (normalized 0-100) in predicting MHI score ranges from smartphone security behaviors.}
\label{fig:MsebisFeatureImportances}
\vspace{-10pt}
\end{figure*}

Our results show that it is possible to predict an individual's mental health status\ignore{ (PHQ9, ISI and IAT)} based on their smartphone security behavioral intentions. We further investigate the role of each individual security attitude on the classification outcome and develop an app to illustrate how these can be monitored in practice on Android which enjoys more than 76\% of the smartphone OS market share~\cite{statistaMarketShare}. Figure~\ref{fig:MsebisFeatureImportances} summarizes our results.

\textbf{Internet addiction:} Based on the best model of prediction (KNN), we found that to observe the smartphone security behaviors in practice one should focus on monitoring \textbf{S5} (see Figure~\ref{fig:fiIaMsebis}). This corresponds to observing if the user deletes any online communications that look suspicious (Table~\ref{table4}). This is not a straightforward behavior to monitor. It would entail monitoring SMSs, emails and posts on online social accounts. On Android monitoring SMSs requires the \texttt{RECEIVE\_SMS} and \texttt{READ\_SMS} permissions are granted. The first allows us to register a broadcast receiver using intent \texttt{android.provider.Telephony.SMS\_RECEIVED} to analyze all incoming SMSs and the latter allows us to traverse all the  received SMSs and determined whether a previously received one had been deleted. However the rest (email and social posts) would require the user providing access to their email and social accounts. Enterprises which have access to privileged operations on mobile OS such as Android and iOS might be able to do that but the same is not true for third party applications. Another important feature for KNN is \textbf{T4}. For brevity we give the intuition: the tracker app creates geolocation spaces corresponding to public spaces and checks the proximity sensor when a sensitive app is in the foreground. The tracker app only requires the \texttt{ACCESS\_FINE\_LOCATION} dangerous permission (accessing the proximity sensor does not require a dangerous permission---android.hardware.sensor.proximity).

\textbf{Depression:} S5 is again the most important feature for KNN (higher recall) with S1 a close second. \textbf{S1} requires monitoring the source of the app when the user performs financial or shopping tasks. This can be achieved using a whitelist of financial and shopping apps (we crawled the 100 top free and paid apps from Google Play's Finance and Shopping categories). To monitor whether any of these are running on the foreground we implemented a background process which periodically uses the \texttt{ActivityManager.getRunningProcesses} Android API to retrieve the current foreground process which is checked against the whitelist. If one chooses Random Forest (higher precision), then the predominantly most important feature to monitor is \textbf{T1} which corresponds to observing whether the user changes their advertising id frequently. This is relatively easy to monitor. Our Android tracker app uses a background thread which in turn uses an \texttt{AdvertisingClient}, the \texttt{getAdvertisingIdInfo()} and \texttt{getID} methods to periodically track changes to the Advertising ID.

\textbf{Insomnia:} If Random Forest is chosen, S5, T1, T3 and T8 are the dominant features to monitor. \textbf{T3} corresponds to tracking changes in the passcode/PIN of the screen lock of the smartphone. We achieve this on Android as follows: we extended the class \texttt{DeviceAdminReceiver} (a BroadcastReveiver subclass on Android) and overrided its \texttt{onPasswordChanged()} method. Note that this is part of the Android Device Administration API and as such it requires the app to request the \texttt{BIND\_DEVICE\_ADMIN} permission. Alternatively a third-party app could leverage the \texttt{isKeyguardSecure()} method of the \texttt{KeygardManager} class to periodically check whether the device is using a secure screen lock method including PIN ,pattern or password, or whether the SIM card is locked.  T8 requires detecting if the user turns off WiFi when not actively being used. We implemented this with a broadcast receiver and a background process. Our broadcast receiver checks for the action \texttt{WifiManager.WIFI\_STATE\_CHANGED\_ACTION} , and if the state of the WiFi is \texttt{WifiManager.WIFI\_STATE\_DISABLING} (i.e., the WiFi is in the process of switching off) we launch a background process which checks the device's network (TCP and UDP) connections as desscribed in Demetriou et al~\cite{demetriou2017hanguard}. Alternatively one could choose Logistic Regression (most important T1, T4) or SVM (T1, T3).

In summary, we are the first to explore and demonstrate the possibility of predicting MHI scores from smartphone security behavior intentions. We further analyzed the importance of each feature on each classification task and how these behaviors can be monitored in practice on the most popular mobile OS.

%% file: discussion.tex
\textbf{Applications of Scale: From a lab to the field.} Our new scale of smartphone security behavior can be employed for various purposes. Researchers may utilize this scale to assess users' intended smartphone security behaviors and investigate how their behaviors change among different populations (similar to Sharif et al.~\cite{sawaya2017self} for SeBIS) or over time with educational interventions. For instance, researchers who are interested in the smartphone malware prevention may use SSBS to explore which smartphone security behaviors may make users more vulnerable to attacks. Researchers can also employ this scale to predict users' behaviors in the field and assess the effectiveness of security interventions (e.g., warnings).

When it comes to the field, SSBS can be applied in different ways. We provide three use cases in different contexts. Firstly, in a healthcare context, a doctor who would like to use health-related apps for treatment or self-management may use this scale to determine if her patients need educational interventions before utilizing the app. Secondly, in a workplace, employers can use this scale to evaluate accidental insider threats on employees' use of smartphone and further implement interventions and measure employees' behavioral changes on smartphone security over time. Thirdly, in an educational context, schools can deploy the scale and assess both teachers and students' smartphone security behaviors while embracing smartphone into online education. Schools may also apply this scale to detect the degree of vulnerability of their students and faculties to potential cyberthreats through smartphones (e.g., cyberbullying, stalking, etc.). Enterprises can also leverage these to design personalized and subject-based cybersecurity educational programs for their employees. 

\textbf{Limitations and Future Work.} In this study, we developed a smartphone security behavior (intentions) scale with high internal consistency and convergent validity. Yet, its predictability of users' \emph{actual} smartphone security behavior has not been confirmed. Therefore, we next plan to conduct a field experiment and test how effective is the scale in predicting users' real behaviors.\ignore{With the predictability of the scale, we can further design a customized security default setting system of smartphone based on users' behavioral tendency.} Also, like the majority of psychometric measurements, SSBS may raise concerns of a self-reported approach. In order to reduce potential social desirability bias, we took several precautionary procedures: being careful about wording questions in a non-judgmental way, making survey anonymous, and keeping the purpose of survey vague. For data cleaning, we excluded unattended responses~\cite{redmiles2017well}. 
In our future work, we will cross-validate the mitigation of social desirability effect by comparing observed data and self-reported data with different populations.

%% file: conclusion.tex
In this work we found that\ignore{the psychological construct of} smartphone security behavior differs from general security behavior\ignore{ measured by SeBIS~\cite{egelman2015scaling}}. Driven by this, we carried out a series of factor analyses to create a Smartphone Security Behavior Scale (SSBS) with 14 questions that load onto two factors: technical (using technical strategies to protect smartphones) and social (being contextually cautious while using smarpthones). Our scale exhibited satisfactory psychometric properties: the full scale and both the subscales have high internal consistency, all items map uniquely on one single component, and no correlation exists between the subscales. We establish convergent validity between SSBS and a well established security behavior measurement\ignore{ SeBIS. These indicate that SSBS is a well-established psychological construct and measurement}. To showcase the application of SSBS we  perform a first-of-its-kind analysis of the effects of popular mental health issues (MHIs)\ignore{(depression, insomnia, internet addiction)} in smartphone security behavior intentions. We found that individuals with high depression indicators are significantly more likely to use technical strategies to protect their smartphones than individuals with low depression indicators; while individuals with low internet addiction indicators are significantly more likely to use social strategies than those with high internet addiction indicators. Lastly we show that smartphone security behavior intentions can be potentially leveraged to profile individuals according to MHIs and illustrate how the most important behaviors for this task can be monitored by a third-party application on the most popular mobile operating system (Android).